\documentclass[10pt,journal]{IEEEtran}
\usepackage{textcomp}
\usepackage[pdftex]{graphicx}
\graphicspath{{../}}
\DeclareGraphicsExtensions{.pdf,.jpeg,.png,.eps}
\usepackage{epstopdf}
\usepackage{algorithm} 
\usepackage{makecell}
\usepackage[frenchb, english]{babel}
\usepackage{csquotes}
\usepackage{cite}
\usepackage{caption}
\usepackage{subcaption}
\usepackage[cmex10]{amsmath}
\usepackage{float,multirow}
\usepackage{amsfonts}
\usepackage[none]{hyphenat}
\usepackage{url}

\usepackage{breakurl}
\usepackage[breaklinks]{hyperref}
\usepackage{soul}
\usepackage{dirtytalk}
\usepackage{hyperref}
\usepackage{comment}
\usepackage[euler]{textgreek}
\usepackage{epsfig,bm,amsmath,amssymb,graphicx,setspace}
\usepackage{tikz}
\usepackage{color}
\usepackage[T1]{fontenc}
\usepackage{algpseudocode}
\usepackage{textgreek}
\usepackage{amsthm}
\usepackage{array}
\usepackage{balance}
\usepackage{enumerate}
\usepackage[utf8]{inputenc}
\usepackage{array}
\usepackage{wrapfig}
\usepackage{multirow}
\usepackage{tabularx}

\usepackage[section]{placeins}

\usepackage{graphicx}
\usepackage{subcaption}
\usepackage[utf8]{inputenc}
\usepackage[export]{adjustbox}
\usepackage{wrapfig}

\usepackage{array}
\usepackage{multirow}
\usepackage{graphicx}
\usepackage[table,xcdraw]{xcolor}

\usepackage{booktabs}  

\usepackage{amssymb}
\usepackage{array}

\renewcommand{\arraystretch}{1.3} 

\def\BibTeX{{\rm B\kern-.05em{\sc i\kern-.025em b}\kern-.08em
        T\kern-.1667em\lower.7ex\hbox{E}\kern-.125emX}}
\usepackage[T1]{fontenc}
\usepackage{multirow}
\usepackage[frenchb]{babel}
\newcolumntype{C}[1]{>{\centering\arraybackslash}m{#1}}
\newcolumntype{P}[1]{>{\centering\arraybackslash}p{#1}}

\begin{document}


  \title{Split Learning-Enabled Framework for Secure and Light-weight Internet of Medical Things Systems}
    
\author{Siva Sai, Manish Prasad, Animesh Bhargava, Vinay Chamola~\IEEEmembership{Senior Member,~IEEE}, Rajkumar Buyya~\IEEEmembership{Fellow,~IEEE}


\thanks{Siva Sai and Rajkumar Buyya are with the Quantum Cloud Computing and Distributed Systems (qCLOUDS) Laboratory, School of Computing and Information Systems, The University of Melbourne, Australia (e-mail: \{sivasaireddy.naga, rbuyya\}@unimelb.edu.au).}
\thanks{Manish Prasad and Vinay Chamola are with the Department of Electrical and Electronics Engineering, BITS-Pilani, Pilani Campus, India 333031 (e-mails: \{p20240903, vinay.chamola\}@pilani.bits-pilani.ac.in).

Animesh Bhargava is with the Department of Computer Science and Information Systems, BITS-Pilani, Pilani Campus, India 333031 (e-mail: h20190545@pilani.bits-pilani.ac.in).

Vinay Chamola is also with APPCAIR, BITS-Pilani, Pilani campus.}


}

\maketitle

\begin{abstract} 
The rapid growth of Internet of Medical Things (IoMT) devices has resulted in significant security risks, particularly the risk of malware attacks on resource-constrained devices. Conventional deep learning methods are impractical due to resource limitations, while Federated Learning (FL) suffers from high communication overhead and vulnerability to non-IID (heterogeneous) data. In this paper, we propose a split learning (SL) based framework for IoT malware detection through image-based classification. By dividing the neural network training between the clients and an edge server, the framework reduces computational burden on resource-constrained clients while ensuring data privacy. We formulate a joint optimization problem that balances computation cost and communication efficiency by using a game-theoretic approach for attaining better training performance. Experimental evaluations show that the proposed framework outperforms popular FL methods in terms of accuracy (+6.35\%), F1-score (+5.03\%), high convergence speed (+14.96\%), and less resource consumption (33.83\%). These results establish the potential of SL as a scalable and secure paradigm for next-generation IoT security.
    \end{abstract}  

\section{Introduction}

Internet of Things (IoT) has emerged as a popular paradigm for connecting vast networks of computing devices, sensors, software, and many more, with enhanced communicative capabilities, automation, and efficiency, thus revolutionizing both industrial and commercial use cases. However, this rapid proliferation of IoT devices has also created many security challenges \cite{schiller2022landscape, hassan2019current}. IoT malware \cite{ngo2020survey} is a type of malicious software specifically designed to target IoT devices. Unlike traditional computer viruses, IoT malware often operates on devices with limited processing power and memory, making it more difficult to detect and remove. Detection of IoT malware using image-based techniques has emerged as a promising approach in cybersecurity. This method leverages visual representations of malware files to identify and classify malicious software targeting IoT devices. By converting malware executable files into grayscale images, unique texture patterns and structural characteristics of the malware code are captured, which are then used as input for specially designed deep Convolutional Neural Networks (CNNs) models, which have proven to be highly effective in extracting discriminative features from these malware images at different abstraction levels; achieving accuracy rates of more than 95\% \cite{jeon2020dynamic}. As opposed to regular deep learning techniques, distributed learning methods are much better suited for IoT devices since these devices often have limited processing power, memory, and energy, and distributed learning methods make better use of all available resources by allowing the computational load to be effectively spread across multiple devices. Federated Learning (FL) \cite{hsu2020privacy,abdel2022efficient} is traditionally used in training IoT-based models. In the context of IoT malware detection, FL enables devices to train detection models collaboratively while preserving data privacy and confidentiality. This approach addresses scalability issues inherent in traditional centralized machine learning methods and reduces communication overhead, which is particularly beneficial in resource-constrained IoT environments \cite{verma2025leveraging}. However, FL faces several limitations, of which the high computational burden on the resource-poor clients is of great concern in IoT scenarios.

\newcommand{\tick}{\checkmark}
\newcommand{\cross}{$\times$}

\newcommand{\half}{%
  \begin{tikzpicture}[baseline=-0.5ex, scale=0.175]
    \draw[thick] (0,0) circle (0.5);
    \fill[black] (0,0) -- (90:0.5) arc (90:270:0.5) -- cycle;
  \end{tikzpicture}%
}

\begin{table*}[t]
\centering
\caption{Comparison with Related Works}
\label{tab:relwork-compact}
\renewcommand{\arraystretch}{1.2}
\small
\begin{tabularx}{\textwidth}{|>{\raggedright\arraybackslash}p{3.2cm}| 
                              >{\raggedright\arraybackslash}X| 
                              >{\centering\arraybackslash}p{0.8cm}| 
                              >{\centering\arraybackslash}p{1.2cm}| 
                              >{\centering\arraybackslash}p{1.2cm}| 
                              >{\centering\arraybackslash}p{1.2cm}| 
                              >{\centering\arraybackslash}p{1cm}|}
\hline
\textbf{Work} & \textbf{Key Contribution} & \textbf{SL} & \textbf{Image\newline based} & \textbf{Comm.\newline eff.} & \textbf{Client\newline compute} & \textbf{Joint\newline opt.} \\
\hline
Asam et al. \cite{asam2022iot} & Modular CNN for IoT malware images & \cross & \tick & \cross & \cross & \cross \\
\hline
Sanchez et al. \cite{rey2022federated}& FL with MLP/AE; centralized-level accuracy & \cross & \cross & \cross & \cross & \cross \\
\hline
Li et al. \cite{li2024split} & Multi-institution SL; 41\% client compute reduction & \tick & \half & \cross & \tick & \cross \\
\hline
Singh et al. \cite{singh2019detailed} & SL efficiency study & \tick & \tick & \half & \half & \cross \\
\hline
\textbf{Proposed Work} & SL for IoMT malware with game-theoretic cost--communication balance & \tick & \tick & \tick & \tick & \tick \\
\hline
\end{tabularx}
\vspace{-2mm}
\begin{flushleft}
\scriptsize
Legend: \tick~present, \half~partial, \cross~not present
\end{flushleft}
\end{table*}

Split Learning (SL) has emerged as a popular alternative in the domain \cite{otoum2022feasibility}, offering several benefits over traditional FL in terms of security and performance. SL allows for more flexible model architectures that can adapt to different privacy requirements and computational constraints of IoT devices. By partitioning the network between layers, SL can offload part of the training process to more powerful servers, alleviating the computational burden on resource-constrained IoT devices. Furthermore, SL has shown superior performance in specific tasks, which could lead to more efficient training of IoT malware detection models. For instance, SL frameworks have accelerated training and improved convergence rates in environmental monitoring scenarios, where scalable and energy-efficient models are critical for deploying water quality and pollution monitoring systems \cite{monitoring1, monitoring2}. SL-based frameworks especially excel in areas where edge nodes operate under limited power supply and are deployed in remote areas, such as forests or hilly regions, for monitoring purposes \cite{battery1, battery2, battery3}.

In this paper, we propose a novel approach that applies SL techniques to enhance IoT malware detection using image-based classification. Our method takes advantage of SL to distribute the computational load between IoT devices and edge servers, using joint optimizations for computation cost for edge devices and communications efficiency between client and server, while maintaining data privacy. By combining image-based IoT malware detection with the proposed SL framework, we aim to develop a more secure, efficient, and privacy-preserving system to protect IoT devices against evolving malware threats. This approach holds promise in addressing the unique challenges posed by the diverse and resource-constrained nature of IoT environments.

The rest of the paper is organized as follows. Section \ref{relwork} describes the works related to the proposed model. In Section \ref{prop}, we describe the proposed methodology, followed by a discussion on performance evaluation in Section \ref{results}. Finally, we conclude the paper in Section \ref{conclusion}.

\section{Related Work}\label{relwork}

\subsection{IoT Malware Detection}
Asam et al. \cite{asam2022iot} proposed the iMDA architecture, which incorporates a modular design featuring edge exploration, multipath dilated convolution, and channel squeezing/boosting in CNN. The architecture was evaluated using a benchmark IoT dataset with image-based malware representations. Their study was limited to static image-based analysis and focused only on ARM-based IoT malware. Rey et al. \cite{rey2022federated} developed a Federated Learning framework using both supervised (multi-layer perceptron) and unsupervised (autoencoder) models and achieved similar accuracy to centralized models (99\% detection rate). However, the model is highly vulnerable to adversarial attacks. Babbar et al. \cite{babbar2024ngmd} proposed a Federated Learning Framework using FedAvg aggregation for distributed attack detection across IoT domains and validated using the BoT-IoT dataset with multiple attack categories (DDoS, reconnaissance). They achieved 99.9\% accuracy after 30 iterations with minimal network delay and memory usage, but faced high communication overhead in large-scale deployments.
Asiri et al. \cite{asiri2025rpfl} introduce a federated learning framework with privacy and reliability enhancements for IoT malware detection by using elliptic curve digital signatures (ECDSA) to verify clients and homomorphic encryption (HE) to protect local model weights and integrating blockchain-based smart contracts \cite{10002899} for decentralized client evaluation and aggregator failure tracking.

\subsection{Split Learning}
Li et al. \cite{li2024split} tested a multi-institutional SL framework on five medical datasets. They compared SL and FL for EHR and imaging data and obtained 96\% agreement with centralized models. 
However, the model was limited to 3-way model splits and incurred high communication latency (around 120ms/round). 
Singh et al. \cite{singh2019detailed} compared the communication efficiencies of split learning and federated learning. The authors demonstrated that the split learning should be favored as the model sizes and the number of clients grow. On the other hand, federated learning becomes more communication-efficient when the number of data samples is limited while keeping the model sizes and number of clients low. A major drawback of this work is that it fails to analyze resource utilization (the number of epochs required for convergence under various scenarios). Table \ref{tab:relwork-compact} compares the proposed model with the related works.

\section{Proposed Methodology}\label{prop}

\begin{figure*}[t]
    \centering
    \includegraphics[width=0.7\linewidth]{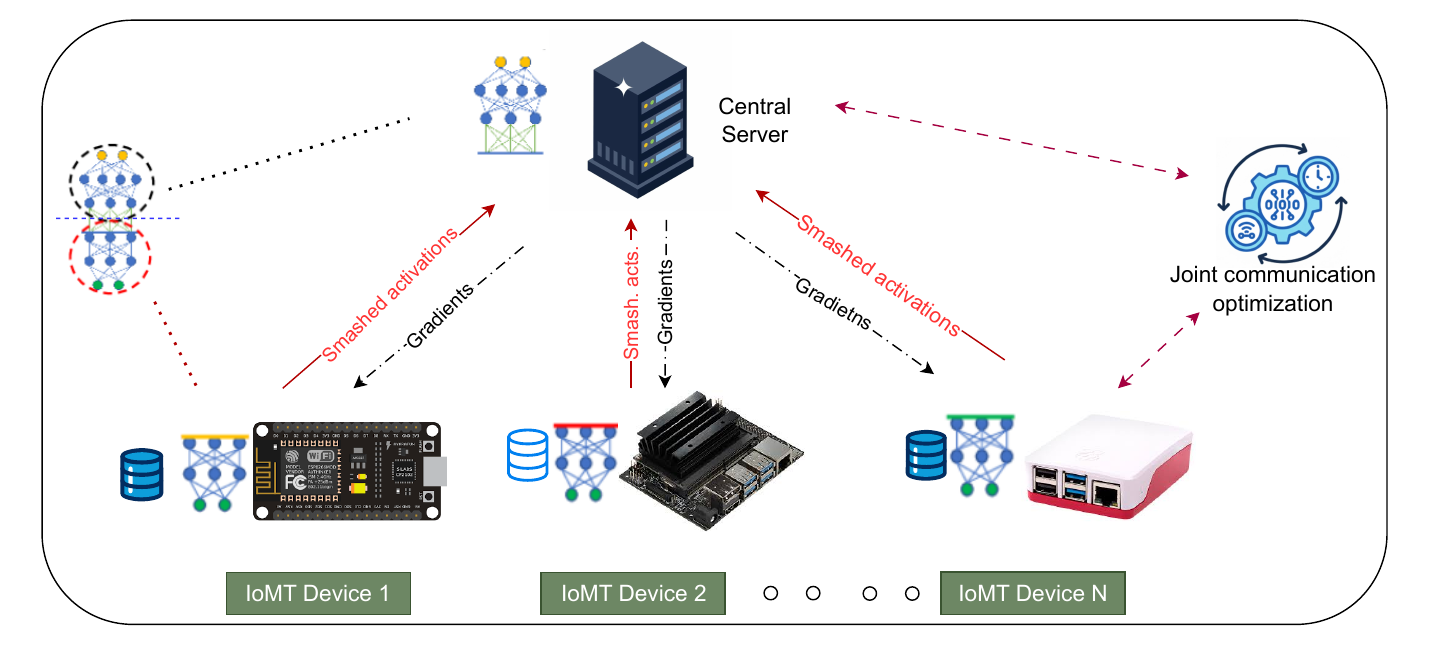}
    \caption{Proposed Split Learning-based Framework for IoMT Malware Detection}
    \label{fig:proposed}
\end{figure*}

In this section, we describe the techniques and methodologies we adopt for implementing the proposed split learning framework (see Figure \ref{fig:proposed}), including the overall workflow, SL algorithms, and optimization techniques. 

\subsection{Overall Workflow}
Initially, IoMT devices and edge servers that want to participate in collaborative training are registered in the system. The split learning paradigm strategically divides the custom Convolutional Neural Network (CNN) between the server and the clients at a predefined cut layer as per the device constraints. To ensure a better distribution of resources, prevent a single point of failure, and improve reliability, the system incorporates dynamic server allocation. Once selected, the server initiates the global model and distributes the client-side portions of the neural network to all participating IoMT devices.

Each training round involves the clients processing their local image-based malware datasets through their portion of the network and transmitting the resulting activations, known as smashed data, to the server. The server continues the forward pass, computes the loss, and performs backpropagation. The newly computed gradients are then sent to the client-side layers and updated their weights accordingly. This process continues to iterate until a predefined stopping condition is reached.

The framework uses a joint strategy for computation and communication efficiency to optimize performance for resource-constrained IoMT devices. Depending on available device resources, cut-layer portions within the CNN are chosen, aiming to minimize client-side computational costs and bandwidth usage. Using stratified sampling and balanced train-validation-test splits improves model generalization despite heterogeneous local datasets. Metrics, including training loss, accuracy, F1 score, processing time, and total computation (TFLOPs), are tracked to compare the split learning approach with traditional federated learning.

\subsection{Learning Framework}

Split learning \cite{vepakomma2018split} has emerged as a distributed and collaborative model training paradigm that addresses many of the issues encountered in federated learning. Algorithm \ref{splitalgo} shows how the model computations take place, including the data transfer between clients and server \cite{10572484}. The model training process in SL involves several steps that are listed below:
\begin{enumerate}
    
    \item 
    Each client processes its private data using its assigned layers of the model, i.e., up to the cut layer. The resulting activations, also called \textit{smashed data}, are securely transmitted to the server.
    
    \item 
    Upon receiving the activations, the server continues the forward pass through the remaining layers of the neural network, generating predictions and computing the training loss.
    
    \item 
    The server performs backpropagation on its portion of the network, obtaining gradients for both its own layers and for the cut layer.
    
    \item 
    The boundary gradients are sent back to the clients. Using these, each client executes backpropagation on its local layers and updates its model parameters.
    
    \item 
    Steps 2--5 are repeated collaboratively across all participating clients for several training rounds. The cycle continues until a convergence condition (e.g., a predefined number of epochs, stable loss, or target accuracy) is met.
\end{enumerate}

\subsection{Joint optimization for computation cost and communication latency}

To increase the model's performance on IoMT edge devices, we adopt a game-theoretic approach to formulate a mathematical model for the joint optimization of computation cost and communication efficiency with minimal compromise in training accuracy. By adapting and extending established formulations from federated learning \cite{10231150} and mobile edge computing, we model a unified optimization problem that formalizes the trade-offs between energy, latency, and accuracy \cite{10908620} in the proposed split learning framework. We consider split learning (SL) over a wireless star topology with a central server and $M$ IoMT devices (clients), indexed by the set $\mathcal{U} \triangleq \{1,\ldots,M\}$. 

Each device $u\in\mathcal{U}$ executes the client-side sub-network up to the split layer and transmits the corresponding smashed activations to the server. We assume frequency-division multiple access (FDMA) for the uplink. As the server's downlink power is typically much larger than client uplink power, downlink latency is negligible and thus ignored.

Let $\mathcal{D}_u=\bigcup_{i=1}^{D_u}\{ \mathbf{x}_{ui},y_{ui}\}$ denote device $u$’s dataset. The SL training aims to minimize the global empirical loss:
\begin{align}
\min_{\mathbf{w}\in\mathbb{R}^{p}} \; F(\mathbf{w}) 
&= \sum_{u=1}^{M} \frac{D_u}{\sum_{j=1}^{M}D_j}\, \ell_u(\mathbf{w}), \\
\ell_u(\mathbf{w}) 
&= \frac{1}{D_u} \sum_{i=1}^{D_u} f_{ui}(\mathbf{w}),
\end{align}
where $f_{ui}(\cdot)$ is the sample loss.

\subsubsection{Uplink Rate, Energy, and Time}
Under FDMA, the achievable uplink rate of device $u$ is
\begin{align}\label{eq:rate_u}
    \varrho_u \;=\; b_u \log_2\!\left(1+\frac{\rho_u \gamma_u}{\sigma^{2} b_u}\right),
\end{align}
where $b_u$ is bandwidth, $\rho_u$ is transmit power, $\gamma_u$ is channel gain, and $\sigma^{2}$ is noise power. Let $\delta_u$ (bits) be the total uplink payload per round. The uplink time and transmission energy per round are:
\begin{equation}
t^{ul}_u = \frac{\delta_u}{\varrho_u}, 
\quad 
\epsilon^{tx}_u = \rho_u \, t^{ul}_u
\end{equation}

\subsubsection{Local Computation Energy and Time}
Let $\phi_u$ be the CPU frequency, $\chi_u$ the cycles per sample, and $\xi$ the switched capacitance. With $L$ local iterations, the local computation energy and time per round are:
\begin{align}\label{eq:e_cmp_t_cmp}
    \epsilon^{\mathrm{cmp}}_{u} = \xi\, L\, \chi_u D_u \phi_u^{2},
    \qquad
    t^{\mathrm{cmp}}_{u} = L\, \frac{\chi_u D_u}{\phi_u}.
\end{align}
Over $G$ global rounds, total energy $\mathcal{E}$ and completion time $T$ are:
\begin{align}
    \mathcal{E} &= G \sum_{u=1}^M \big(\epsilon^{\mathrm{tx}}_{u} + \epsilon^{\mathrm{cmp}}_{u}\big), \label{eq:total_energy} \\
    T &= G \max_{u\in\mathcal{U}}\big\{ t^{\mathrm{cmp}}_{u} + t^{\mathrm{ul}}_{u} \big\}. \label{eq:total_time}
\end{align}

\begin{algorithm}[!t]
\caption{Resource Allocation for SL (Energy--Latency Trade-off)}
\label{alg:ra_sl}
\small
\begin{algorithmic}[1]
\State Initialize feasible $(\{\rho_u^{(0)}\},\{b_u^{(0)}\})$ and $\{\phi_u^{(0)}\}$; set $k\!\leftarrow\!0$.
\Repeat
\State \textbf{(Subproblem A)} Fix $\varrho_u \!=\! \mathcal{G}_u(\rho_u^{(k)},b_u^{(k)})$ and solve \eqref{prob:subA_dual} for $\{\ell_u^{\star}\}$; set $\phi_u^{(k+1)} \leftarrow \mathrm{clip}\!\left(\sqrt[3]{\ell_u^{\star}/(2\alpha G \xi)}, \phi_u^{\min}, \phi_u^{\max}\right)$ and compute the tightest $\vartheta^{(k+1)}$ from \eqref{cons:time_cap}.
\State \textbf{(Subproblem B)} With $\phi_u^{(k+1)}$ and $\vartheta^{(k+1)}$, set $\underline{\varrho}_u$ via \eqref{eq:rate_min}. Initialize $(\upsilon_u,\zeta_u)$ from $(\rho_u^{(k)},b_u^{(k)})$.
\Repeat 
    \State Solve \eqref{prob:subB_subtractive} using KKT \eqref{eq:kkt_rho}--\eqref{eq:kkt_b} and \eqref{eq:bu_star} to obtain $(\rho_u,b_u)$.
    \State Update $\upsilon_u \!\leftarrow\! \alpha G/\mathcal{G}_u(\rho_u,b_u)$,~ $\zeta_u \!\leftarrow\! \rho_u\delta_u/\mathcal{G}_u(\rho_u,b_u)$.
\Until{Newton-like residual $\|\Phi(\zeta,\upsilon)\|$ below tolerance}
\State Set $(\rho_u^{(k+1)},b_u^{(k+1)})\leftarrow(\rho_u,b_u)$.
\State $k\leftarrow k+1$.
\Until{relative change in $(\{\rho_u\},\{b_u\},\{\phi_u\})$ below threshold}
\end{algorithmic}
\end{algorithm}

\subsubsection{Optimization Problem} \label{subsec:opt_problem_sl}
We jointly allocate transmit powers $\{\rho_u\}$, bandwidths $\{b_u\}$ and CPU frequencies $\{\phi_u\}$ to balance energy and time, using a weight $\alpha\in[0,1]$.
\begin{subequations}\label{prob:main}
\begin{align}
\min_{\{\rho_u,\phi_u,b_u\},~\vartheta}~~ 
& \alpha G \sum_{u=1}^M \!\left(\rho_u \frac{\delta_u}{\varrho_u} + \xi L \chi_u D_u \phi_u^2\right) +(1-\alpha) G \vartheta \\
\text{s.t.}\quad 
& t^{\mathrm{cmp}}_{u} + t^{\mathrm{ul}}_{u} \le \vartheta,~~\forall u\in\mathcal{U}, \label{cons:time_cap}\\
& \rho_u^{\min} \le \rho_u \le \rho_u^{\max},~~ \phi_u^{\min} \le \phi_u \le \phi_u^{\max},~~\forall u, \\
& \sum_{u=1}^{M} b_u \le \bar{b}, \quad b_u \ge 0,~~\forall u,
\end{align}
\end{subequations}
where $\vartheta$ is an auxiliary variable for per-round latency. Constraint \eqref{cons:time_cap} is equivalent to:
\begin{align}\label{eq:rate_min}
    \varrho_u \;\ge\; \underline{\varrho}_u \triangleq \frac{\delta_u}{\vartheta - \frac{L\chi_u D_u}{\phi_u}}, \qquad \forall u\in\mathcal{U}.
\end{align}
Problem \eqref{prob:main} is non-convex, so we use problem decomposition.

\begin{algorithm}[]
\caption{Proposed Split Learning Procedure}
\label{splitalgo}
\begin{algorithmic}[1]
    \State \textbf{Setup:}
    \State The server initializes the global model $M$
    \State
    
    \State \textbf{On the Server ($\mathbb{S}$):}
    \While{stopping condition not satisfied}
        \State Collect gradient or activation updates $\mathbb{U}$ from participating clients
        \For{each client $i$}
            \State Aggregate local contributions $m$ into the global model $M$
        \EndFor
        \State Broadcast the refined global model $M'$ back to all clients
    \EndWhile
    \State \textbf{Terminate}
    
    \State
    
    \State \textbf{On Client $i$ ($\mathbb{C}_i$):}
    \While{training not converged}
        \State Compute local gradients $\nabla_{\theta_i} L_i$ using dataset $\mathbb{D}_i$ and parameters $\theta_i$
        \State Forward the computed gradients $\nabla_{\theta_i} L_i$ to the server
        \State Update local parameters by incorporating the received global model $M'$
    \EndWhile
    \State \textbf{Terminate}
\end{algorithmic}
\end{algorithm}

\subsubsection{Problem Decomposition}
We separate \eqref{prob:main} into (i) frequency--latency allocation over $(\{\phi_u\},\vartheta)$ and (ii) radio allocation over $(\{\rho_u,b_u\})$.
\paragraph{Subproblem A (CPU \& latency).} For fixed $\varrho_u$, this subproblem is convex.
\begin{subequations}\label{prob:subA}
\begin{align}
\min_{\{\phi_u\},\vartheta}~~ & \alpha G \sum_{u=1}^{M} \xi L \chi_u D_u \phi_u^2 + (1-\alpha) G \vartheta \\
\text{s.t.}\quad & L\frac{\chi_u D_u}{\phi_u} + \frac{\delta_u}{\varrho_u} \le \vartheta,~~ \phi_u^{\min} \le \phi_u \le \phi_u^{\max},~~\forall u.
\end{align}
\end{subequations}
Applying KKT conditions yields the optimal CPU frequency:
\begin{align}
\phi_u^{\star} = \sqrt[3]{\frac{\ell_u}{2\alpha G \xi}}, \label{eq:phi_star}
\end{align}
where $\ell_u$ are Lagrange multipliers. The dual problem is a convex resource-splitting program:
\begin{subequations}\label{prob:subA_dual}
\begin{align}
\max_{\ell_u\ge 0}~~ 
& \sum_{u=1}^{M} \big((2^{-\frac{2}{3}} + 2^{\frac{1}{3}})\, L (\alpha\xi G)^{\frac{1}{3}} \chi_u D_u\, \ell_u^{\frac{2}{3}} + \tfrac{\delta_u}{\varrho_u} \ell_u \big) \\
\text{s.t.}\quad & \sum_{u=1}^{M}\ell_u = (1-\alpha)G.
\end{align}
\end{subequations}

\paragraph{Subproblem B (power \& bandwidth).} Define $\mathcal{G}_u(\rho_u,b_u) \triangleq b_u \log_2(1+\frac{\rho_u \gamma_u}{\sigma^2 b_u})$. The radio subproblem is:
\begin{subequations}\label{prob:subB_epi}
\begin{align}
\min_{\{\rho_u,b_u,\zeta_u\}}~~ & \alpha G \sum_{u=1}^{M} \zeta_u \\
\text{s.t.}\quad 
& \rho_u^{\min}\le \rho_u \le \rho_u^{\max},~~ \sum_{u=1}^{M} b_u \le \bar{b}, \\
& \mathcal{G}_u(\rho_u,b_u) \ge \underline{\varrho}_u,~~ \rho_u\delta_u - \zeta_u \mathcal{G}_u(\rho_u,b_u) \le 0,\forall u.
\end{align}
\end{subequations}
This can be solved via an equivalent subtractive-form program by finding $\{\upsilon_u^{\star}\}$ such that solving for fixed $(\upsilon_u,\zeta_u)$:
\begin{subequations}\label{prob:subB_subtractive}
\begin{align}
\min_{\{\rho_u,b_u\}}~~ & \sum_{u=1}^{M} \upsilon_u\big(\rho_u \delta_u - \zeta_u \mathcal{G}_u(\rho_u,b_u)\big) \\
\text{s.t.}\quad & \rho_u^{\min}\le \rho_u \le \rho_u^{\max},~~ \sum b_u \le \bar{b},~~ \mathcal{G}_u \ge \underline{\varrho}_u,
\end{align}
\end{subequations}
and then updating $\upsilon_u \leftarrow \frac{\alpha G}{\mathcal{G}_u(\rho_u,b_u)}$ and $\zeta_u \leftarrow \frac{\rho_u \delta_u}{\mathcal{G}_u(\rho_u,b_u)}$ drives the solution. The KKT conditions for \eqref{prob:subB_subtractive} yield closed-form characterizations. Let $\omega_u \triangleq \frac{\rho_u \gamma_u}{\sigma^{2} b_u}$. Optimality implies:
\begin{align}
&\upsilon_u\!\left(\delta_u - \frac{\zeta_u \gamma_u}{\sigma^{2}(1+\omega_u)\ln 2}\right) - \frac{\theta_u \gamma_u}{\sigma^{2}(1+\omega_u)\ln 2} = 0, \label{eq:kkt_rho}\\
&-(\upsilon_u \zeta_u + \theta_u) \log_2(1+\omega_u) \nonumber \\ 
&+ \frac{(\upsilon_u \zeta_u + \theta_u)\rho_u \gamma_u}{(1+\omega_u)\ln 2\, \sigma^{2} b_u} + \mu = 0, \label{eq:kkt_b}\\ 
&\theta_u \Big(b_u \log_2(1+\omega_u) - \underline{\varrho}_u \Big) = 0, \label{eq:kkt_theta}\\
&\mu\!\left(\sum_{j=1}^{M} b_j - \bar{b}\right)=0. \label{eq:kkt_mu}
\end{align}
Solving the above yields:
\begin{align}
b_u^{\star} &=
\begin{cases}
\displaystyle \frac{\underline{\varrho}_u}{\log_2(1+\Lambda_u)}, & \text{if } \theta_u \neq 0,\\[8pt]
\text{solution of \eqref{prob:subB_subtractive} with }\theta_u=0, & \text{otherwise,}
\end{cases} \label{eq:bu_star}\\
\rho_u^{\star} &= \min\!\left\{\rho_u^{\max},\;\max\!\big\{\Gamma(b_u),\, \rho_u^{\min}\big\}\right\}, \label{eq:rho_star}\\
\Gamma(b_u) &\triangleq \left(\frac{(\upsilon_u\zeta_u + \theta_u)\gamma_u}{\sigma^{2}\delta_u \upsilon_u \ln 2} - 1\right)\frac{\sigma^{2} b_u}{\gamma_u}, \label{eq:Gamma}\\
\Lambda_u &\triangleq \frac{(\upsilon_u\zeta_u + \theta_u)\gamma_u}{\sigma^{2}\delta_u \upsilon_u \ln 2}. \label{eq:Lambda}
\end{align}

\subsection{Alternating Optimization and Convergence Criterion}
We solve \eqref{prob:main} by alternating between Subproblem A and Subproblem B as shown in Algorithm \ref{alg:ra_sl}.

\noindent
In the inner loop, we employ a diagonal Newton step with $\Phi\!=\![\Phi_1^\top,\Phi_2^\top]^\top\!\in\mathbb{R}^{2M}$,
\begin{align}
\Phi_1(\zeta) 
&= \bigl[
  -\rho_1 \delta_1 + \zeta_1 \mathcal{G}_1(\rho_1,b_1),~
  \ldots, \notag\\
&\hspace{3.8em}
  -\rho_M \delta_M + \zeta_M \mathcal{G}_M(\rho_M,b_M)
\bigr]^{\top}, \label{eq:Phi1}\\[0.5em]
\Phi_2(\upsilon) 
&= \bigl[
  -\alpha G + \upsilon_1 \mathcal{G}_1(\rho_1,b_1),~
  \ldots, \notag\\
&\hspace{3.8em}
  -\alpha G + \upsilon_M \mathcal{G}_M(\rho_M,b_M)
\bigr]^{\top}. \label{eq:Phi2}
\end{align}
whose Jacobians are diagonal with entries $\mathcal{G}_u(\rho_u,b_u)$, ensuring efficient updates. The alternating optimization method guarantees convergence because each subproblem is convex, each update weakly decreases the global objective, and the objective is bounded below by 0.


Unlike federated learning (FL), split learning (SL) partitions the network such that clients transmit \emph{smashed activations} rather than full parameter vectors. This distinction is reflected in our formulation through two key parameters: (i)~$\delta_u$, representing the SL payload per global round, and (ii)~$\chi_u$, capturing the number of client-side CPU cycles per data sample. The optimization problem \eqref{prob:main} therefore balances the computation energy ($\propto\phi_u^2$) against communication latency (enforced via $\underline{\varrho}_u$). The weight $\alpha\in[0,1]$ tunes this balance. In energy-limited scenarios, setting $\alpha\!\to\!1$ prioritizes minimizing total energy consumption. Conversely, in latency-critical use cases, setting $\alpha\!\to\!0$ emphasizes minimizing overall completion time. Thus, our formulation provides a flexible framework to adapt SL resource allocation to diverse system objectives and hardware constraints.

\section{Performance Evaluation}\label{results}




\begin{figure*}[!ht]
 \begin{subfigure}{0.32\textwidth}
     \includegraphics[width=1\textwidth]{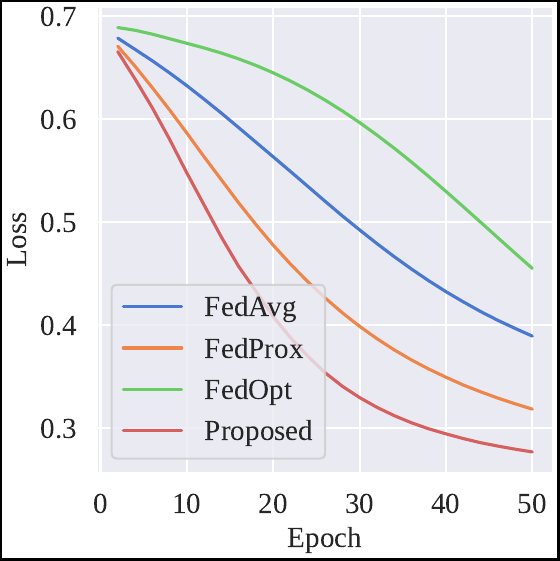}
     \caption{Loss vs Epoch}
     \label{fig:epoch_loss}
 \end{subfigure}
 \hfill
 \begin{subfigure}{0.32\textwidth}
     \includegraphics[width=1\textwidth]{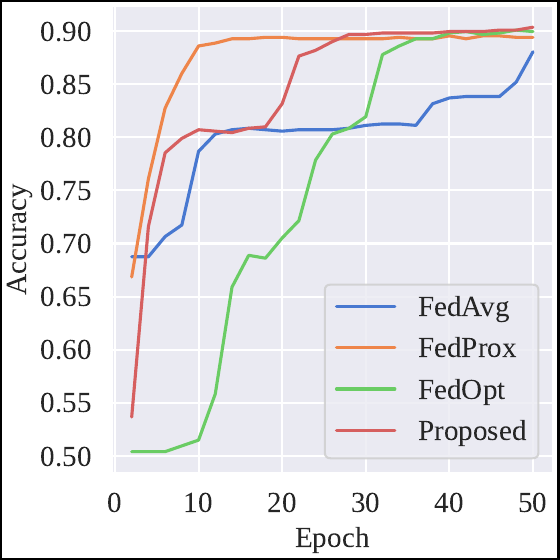}
     \caption{Accuracy vs Epoch}
     \label{fig:epoch_acc}
 \end{subfigure}
 \hfill
 \begin{subfigure}{0.32\textwidth}
     \includegraphics[width=1\textwidth]{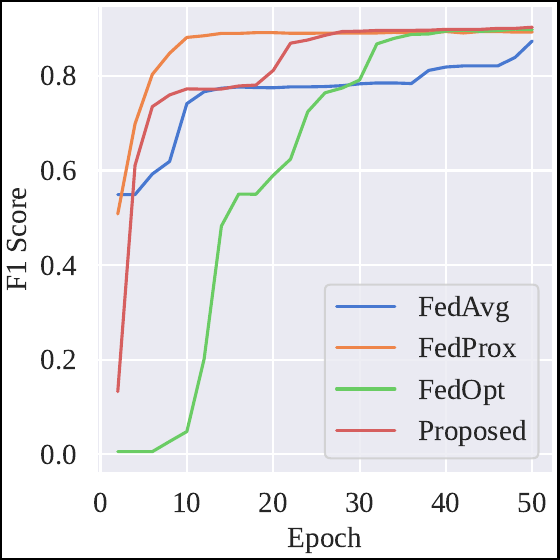}
     \caption{F1-Score vs Epoch}
     \label{fig:epoch_f1}
 \end{subfigure}
 
 \begin{subfigure}{0.32\textwidth}
     \includegraphics[width=1\textwidth]{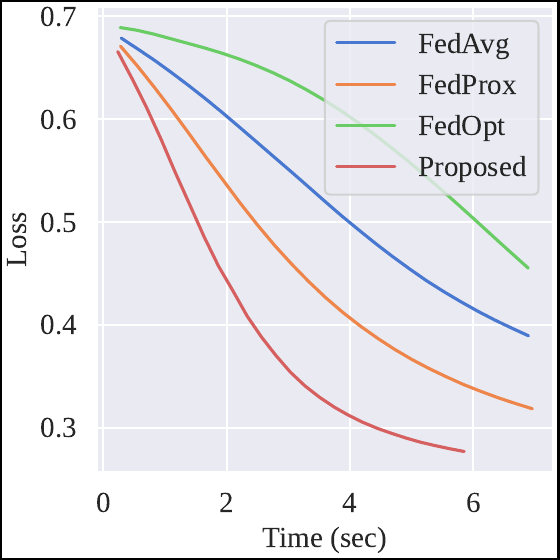}
     \caption{Loss vs Time (sec)}
     \label{fig:time_loss}
 \end{subfigure}
 \hfill
 \begin{subfigure}{0.32\textwidth}
     \includegraphics[width=1\textwidth]{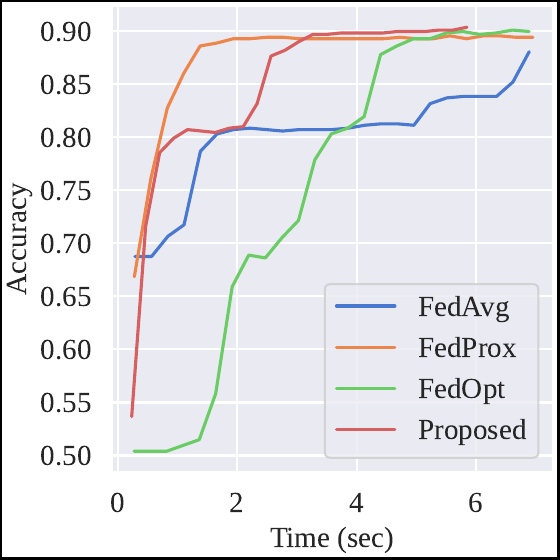}
     \caption{Accuracy vs Time (sec)}
     \label{fig:time_ac}
 \end{subfigure}
 \hfill
 \begin{subfigure}{0.32\textwidth}
     \includegraphics[width=1\textwidth]{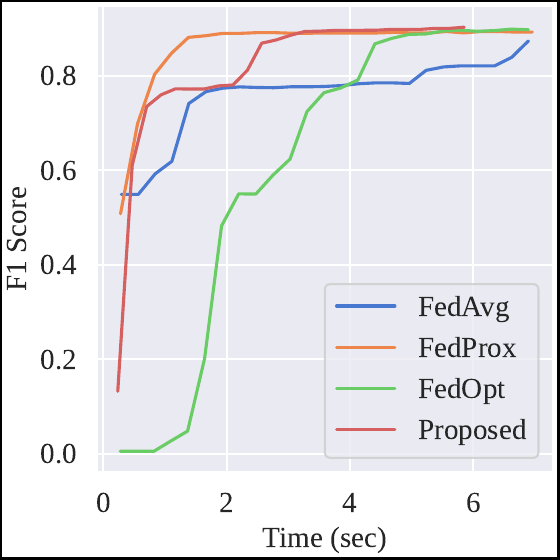}
     \caption{F1-score vs Time (sec)}
     \label{fig:time_f1}
 \end{subfigure}
 
 \begin{subfigure}{0.32\textwidth}
     \includegraphics[width=1\textwidth]{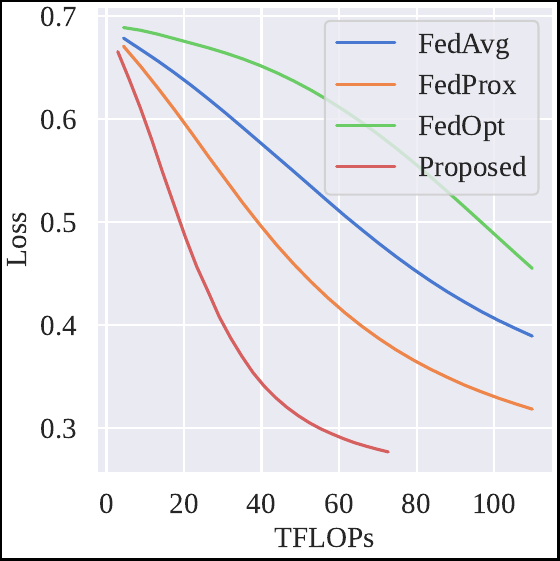}
     \caption{Loss vs TFLOPs}
     \label{fig:tflop_loss}
 \end{subfigure}
 \hfill
 \begin{subfigure}{0.32\textwidth}
     \includegraphics[width=1\textwidth]{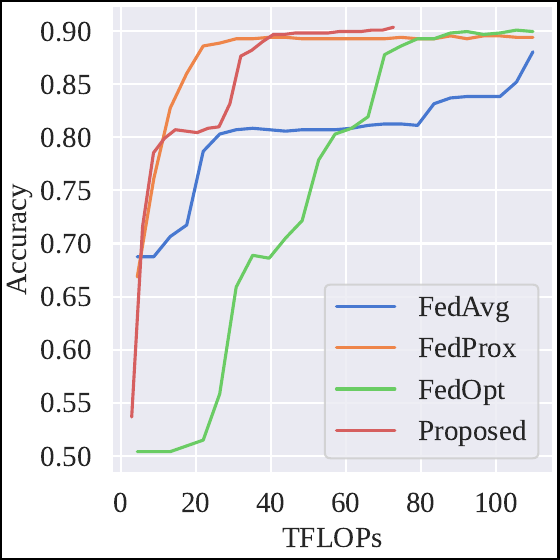}
     \caption{Accuracy vs TFLOPs}
     \label{fig:tflop_acc}
 \end{subfigure}
 \hfill
 \begin{subfigure}{0.32\textwidth}
     \includegraphics[width=1\textwidth]{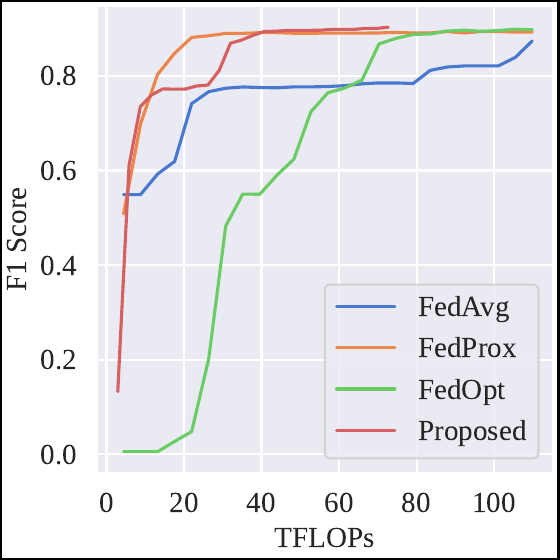}
     \caption{F1-score vs TFLOPs}
     \label{fig:tflop_f1}
 \end{subfigure}
 
 \caption{\textcolor{black}{Comparing loss, accuracy, and F1-score of the proposed SL framework with the baseline FL models across epochs, average client processing times, and total client processing TFLOPs}}
 \label{all_validation_metrics}
\end{figure*}

\begin{table*}
\centering
\caption{Testing performance metrics comparison of the proposed SL framework with baseline FL models}
\label{tab:testing_performance}
\begin{tabular}{lcccc}
\hline
\textbf{Method} & \textbf{Accuracy (\%)} & \textbf{Precision (\%)} & \textbf{Recall (\%)} & \textbf{F1-Score (\%)} \\ \hline
FedAvg          & 87.64                  & 93.95                   & 80.38                & 86.64                  \\
FedProx         & 89.67                  & 91.69                   & 87.19                & 89.39                  \\
FedOpt          & 89.27                  & 92.60                   & 85.29                & 88.79                  \\
Proposed        & 90.49                  & 92.07                   & 88.56                & 90.28                  \\ \hline
\end{tabular}
\end{table*}


\begin{table}[h]
\centering
\caption{Comparison of average client training processing time and resource consumption across different federated learning methods}
\label{tab:time_tflops_comparison}
\begin{tabular}{lcc}
\hline
\textbf{Method} & \textbf{Time (sec)} & \textbf{TFLOPs} \\ \hline
FedAvg          & 6.89                & 109.87          \\
FedProx         & 6.95                & 109.87          \\
FedOpt          & 6.89                & 109.87          \\
Proposed        & 5.85                & 72.70           \\ \hline
\end{tabular}
\end{table}

\begin{figure*}

 \begin{subfigure}{0.32\textwidth}
     \includegraphics[width=1\textwidth]{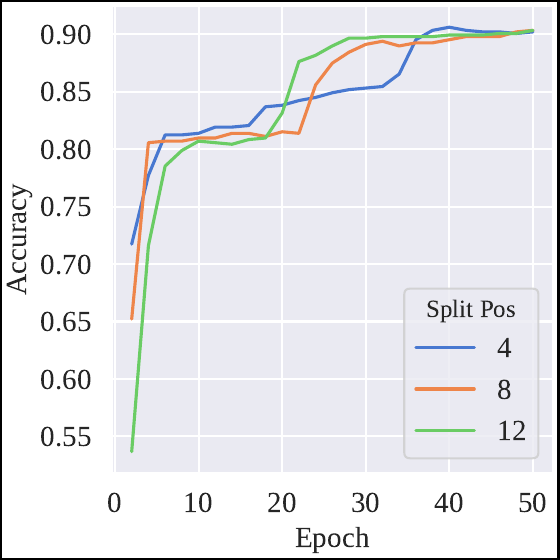}
     \caption{Accuracy vs Epoch (sec)}
     \label{fig:cut_acc_epoch}
 \end{subfigure}
 \hfill
 \begin{subfigure}{0.32\textwidth}
     \includegraphics[width=1\textwidth]{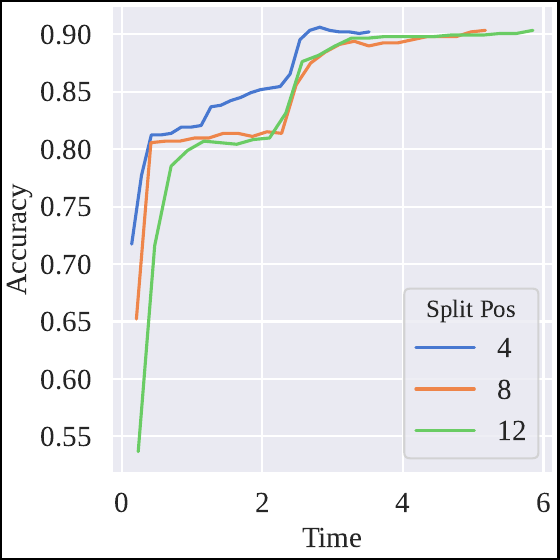}
     \caption{Accuracy vs Time (sec)}
     \label{fig:cut_acc_time}
 \end{subfigure}
 \hfill
 \begin{subfigure}{0.32\textwidth}
     \includegraphics[width=1\textwidth]{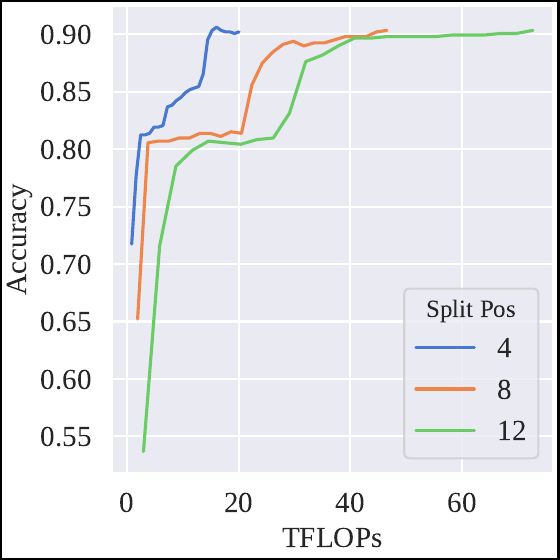}
     \caption{Accuracy vs TFLOPs}
     \label{fig:cut_acc_tflop}
 \end{subfigure}

  \begin{subfigure}{0.32\textwidth}
     \includegraphics[width=1\textwidth]{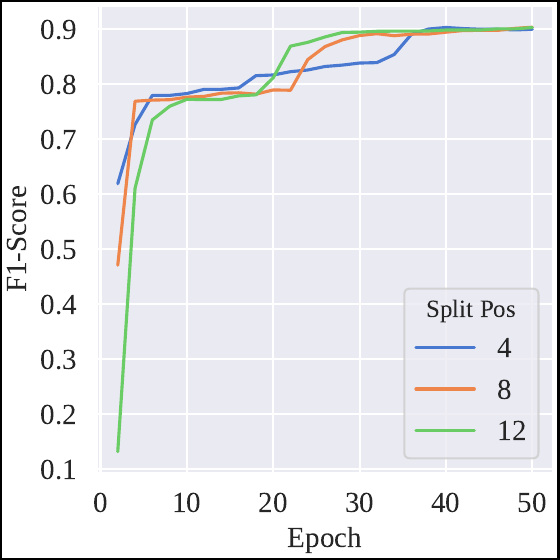}
     \caption{F1-score vs Epoch}
     \label{fig:cut_f1_epoch}
 \end{subfigure}
 \hfill
 \begin{subfigure}{0.32\textwidth}
     \includegraphics[width=1\textwidth]{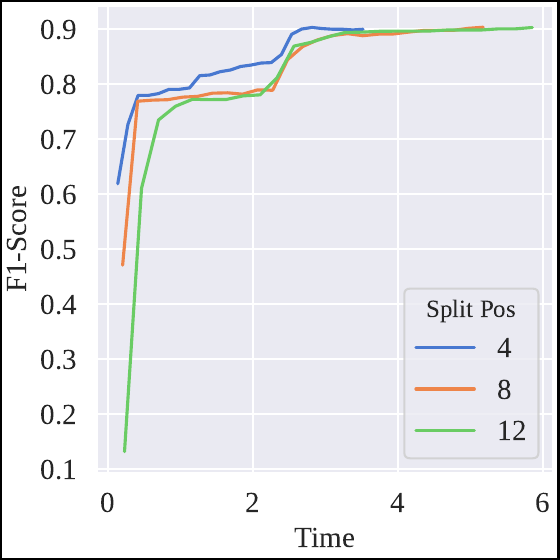}
     \caption{F1 Score vs Time (sec)}
     \label{fig:cut_f1_time}
 \end{subfigure}
 \hfill
 \begin{subfigure}{0.32\textwidth}
     \includegraphics[width=1\textwidth]{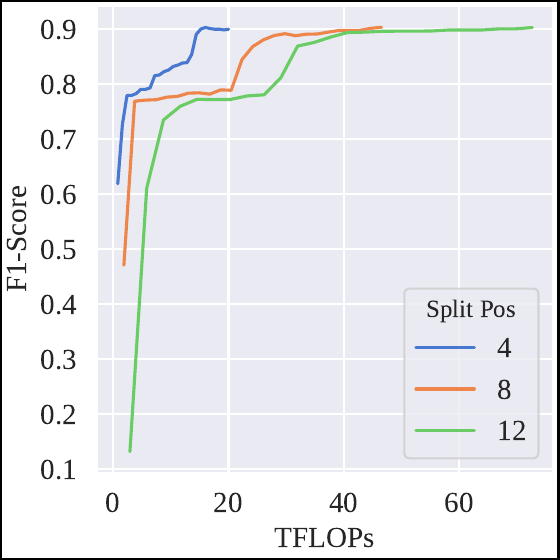}
     \caption{F1-score vs TFLOPs}
     \label{fig:cut_f1_tflop}
 \end{subfigure}
 
 \caption{\textcolor{black}{Comparing accuracy and F1-scores of the proposed SL framework for different split positions across epochs, average client processing times, and total client processing TFLOPs}}
 \label{fig:cut_exp_comp}

\end{figure*}

\begin{figure*}

 \begin{subfigure}{0.32\textwidth}
     \includegraphics[width=1\textwidth]{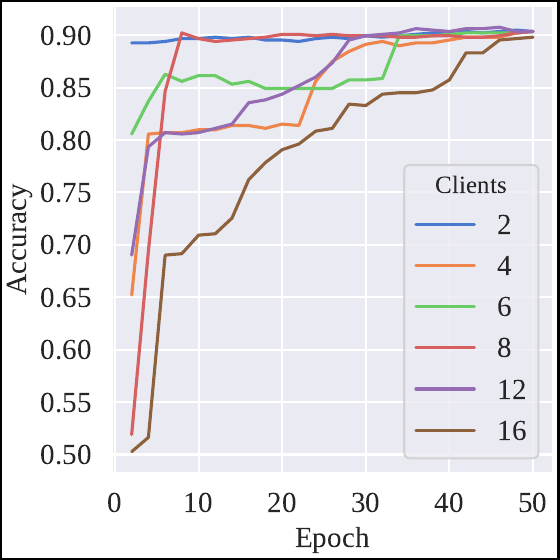}
     \caption{Accuracy vs Epochs}
     \label{fig:client_acc_epoch}
 \end{subfigure}
 \hfill
 \begin{subfigure}{0.32\textwidth}
     \includegraphics[width=1\textwidth]{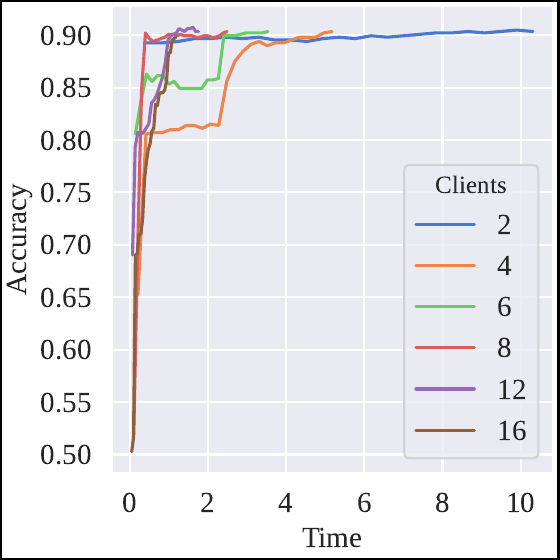}
     \caption{Accuracy vs Time (sec)}
     \label{fig:client_acc_time}
 \end{subfigure}
 \hfill
 \begin{subfigure}{0.32\textwidth}
     \includegraphics[width=1\textwidth]{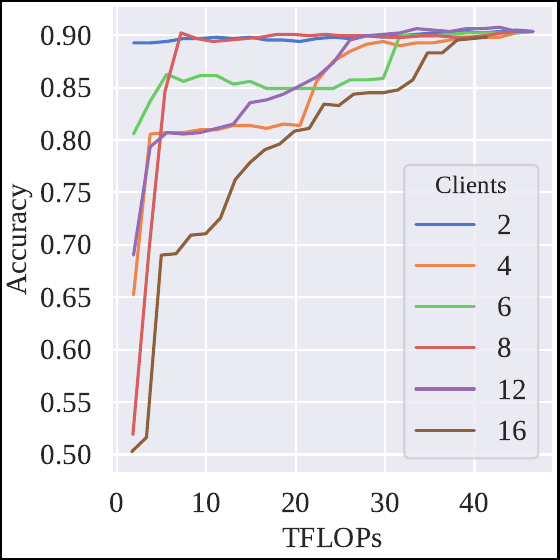}
     \caption{Accuracy vs TFLOPs}
     \label{fig:client_acc_tflop}
 \end{subfigure}

  \begin{subfigure}{0.32\textwidth}
     \includegraphics[width=1\textwidth]{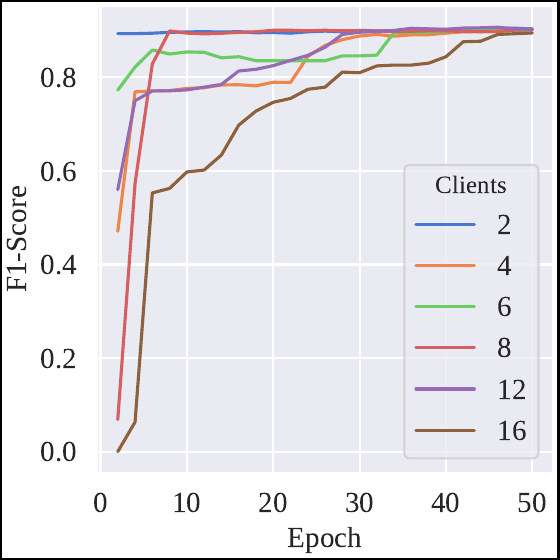}
     \caption{F1-score vs Epochs}
     \label{fig:client_f1_epoch}
 \end{subfigure}
 \hfill
 \begin{subfigure}{0.32\textwidth}
     \includegraphics[width=1\textwidth]{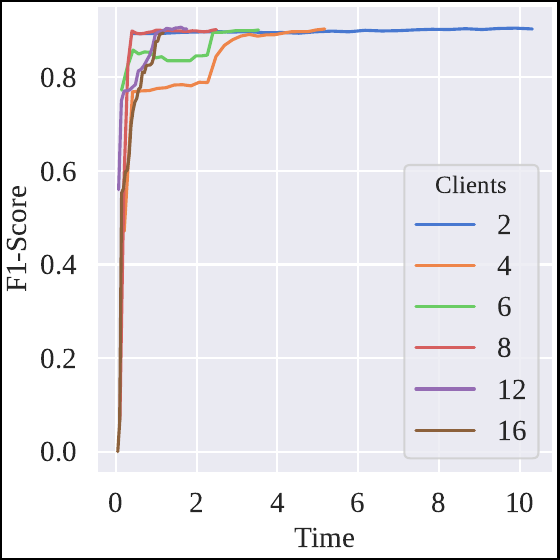}
     \caption{F1-score vs Time (sec)}
     \label{fig:client_f1_time}
 \end{subfigure}
 \hfill
 \begin{subfigure}{0.32\textwidth}
     \includegraphics[width=1\textwidth]{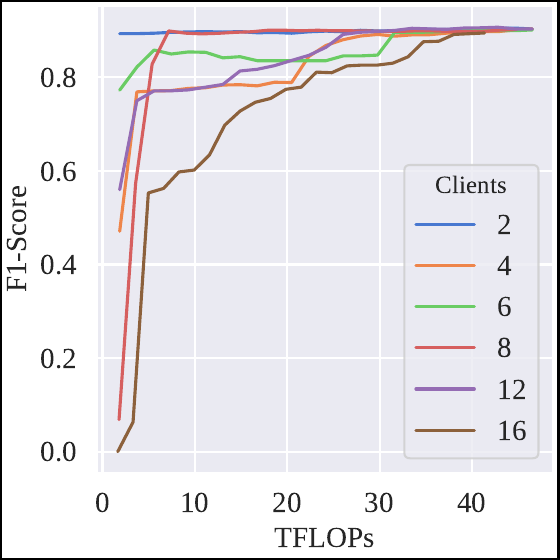}
     \caption{F1-score vs TFLOPs}
     \label{fig:client_f1_tflop}
 \end{subfigure}

 \caption{\textcolor{black}{Comparing performance of proposed split learning framework based on the number of clients participating in the feed forward network}}
 \label{client_comp}

\end{figure*}

In this section, we present several experimental results and analyses related to the proposed model. We also describe the dataset and experimentation tools employed in our work. 

\noindent\textbf{Dataset:} We have used the IoT malware image dataset \footnote{\url{https://www.kaggle.com/datasets/anaselmasry/iot-malware/data}} for carrying out the experimental analysis in our work. The dataset contains 17219 images of dimensions $299\times299$ across three channels. The dataset has a binary classification structure with malware and benign being the two corresponding classes. The dataset is highly imbalanced, with the malignant samples comprising only 14.44\% of the entire dataset. To deal with this imbalance, we down-sample the majority benign class and apply stratified sampling to ensure a balanced train-validation-test split.

\noindent\textbf{Configuration of ML model:} The train-validation-test split we have used is 70-15-15\%. After creating the splits, we perform pixel-wise normalization computation for the train split and apply appropriate transforms to architect a rich dataset for training. For the experiments comparing the performance of the proposed SL model and existing techniques, we use a custom-defined Convolution Neural Network (CNN) with 18 hidden layers constructed with the help of PyTorch and related libraries. This gives us a CNN with four sequential blocks with four layers each - Conv2D, ReLU, 0.5-dropout, and half-resolution MaxPool2d, and a fully-connected block with linear, ReLU, and a dropout layer for regularization. The convolution layers use an 8-16-32-64 feature maps sequence with kernel, stride, and padding of size 3, 1, 1, respectively. The fully connected linear layer uses 512 features for binary classification. The implementation of the FL algorithm that we use for our experiments uses this custom-defined CNN model for both clients and the server. 

\noindent\textbf{Configuration of SL framework:} For the proposed split learning implementation, the four sequential blocks provide us with three possible cut layers to split the model at layer positions 4th, 8th, and 12th, respectively. The client models in this case simulate the decentralized IoMT environments. 
For the training process, we have used the cross-entropy function for loss computation and the Adam optimizer for gradient descent optimization. Both the FL and SL models are trained for an equivalent of 50 iterations (25 rounds with two epochs for each client) with a batch size of 32 samples per iteration. We use client counts between 2 and 16 for both models to gather a comprehensive report and compare their performance. We track key training parameters for monitoring training progress and performance comparison, including training loss, validation accuracy, F1-score, precision, and recall, processing time for backward and forward passes for client and server, and computation cost in Tera Floating-point Operations (TFLOPs) for each client. We use several Python-based libraries like PyTorch, scikit-learn, numpy, pandas, seaborn, and torchprofile to conduct the experiments. All the experiments in the proposed work are carried out using T4 GPU environments on Google Colaboratory\footnote{\url{https://colab.research.google.com/}}.

\noindent\textbf{Configuration of joint optimization framework: }To validate the mathematical model for resource allocation, we simulated the framework using parameters consistent with established research in wireless federated learning. The simulation environment models a star network consisting of a central base station and 50 devices. These devices are distributed uniformly within a circular area of 500m x 500m. The wireless channel is modeled with a path loss of 128.1 + 37.6 log(distance) (km) and a Gaussian noise power spectral density of -174 dBm/Hz. For the learning process, the number of global aggregation rounds is set to 400, with each round comprising 10 local iterations. Each device is assigned 500 data samples, and the upload data size per device is 28.1 kbits. Device hardware characteristics include a maximum CPU frequency of 2 GHz, a maximum transmission power of 12 dBm, and a total available system bandwidth of 20 MHz. We utilized Python-based numerical optimization library, CVXPY 1.5.2 \footnote{\url{https://github.com/cvxpy/cvxpy}}, for the purpose, leveraging the ECOS and SCS solvers for convex and conic constraints. Additionally, we used NumPy for the vectorized computations, SciPy for nonlinear optimization and root-finding, and SymPy for symbolic derivation of analytical gradients and KKT conditions. Iterative convergence was monitored through a residual norm tolerance of $10^-4$.

\subsection{Comparing proposed model with the FL baselines}

Figure \ref{all_validation_metrics} compares the training performance of the proposed SL model with the baseline FL models, namely FedAvg \cite{mcmahan2017communication}, FedProx\cite{li2020federated}, and FedOpt\cite{reddi2020adaptive}, across 50 epochs. The results are shown for federated models with four clients (FdA-4, FdP-4, FdO-4) and SL with four clients and cut position 1, 12th layer (SpL-4-1). The number of iterations is represented on the X-axis, while the corresponding performance metric is represented on the Y-axis. Figure \ref{fig:epoch_loss} shows the average training loss across clients for all the federated and split models, while Figures \ref{fig:epoch_acc} and \ref{fig:epoch_f1} showcase the validation accuracy and F1-scores at the end of each round for all the models. Figure \ref{fig:epoch_loss} shows that the proposed SpL-4-1 model is much better at fitting the training data in a decentralized way than federated models, both in terms of speed and quality. The convergence speeds follow the trend: $\text{SpL4} > \text{FdP-4} > \text{FdA-4} > \text{FdO-4}$

Figure \ref{fig:epoch_acc} indicates that SpL-4-1 consistently outperforms FdA-4 in terms of accuracy as well, while the end performance remains similar compared to FdP-4 and FdO-4. While the accuracy curve for FdP-4 exhibits a steep initial rise, signifying faster convergence, SpL-4-1 ultimately maintains a rapid and consistent learning trajectory, stabilizing at a higher accuracy level of 90.33\% compared to FdA-4, FdP-4, and FdO-4 with 88.04\%, 89.4\%, and 89.94\% accuracy at the end of equivalent epochs, respectively. The F1-score comparison in Figure \ref{fig:epoch_f1} further highlights these performance dynamics. Notably, FdO-4 exhibits a significant "cold start" period with a zero F1-score, indicating initial difficulty in generalizing. By the final epoch, SpL-4-1, FdP-4, and FdO-4 converge to a similar, high F1-score, significantly outperforming the baseline FdA-4. These results demonstrate that the proposed SL framework enables more effective model training and improved generalization compared to standard FL approaches. 
Table \ref{tab:testing_performance} presents the test result metrics comparing all the stated models. These results suggest that SL enables more effective model training and improved generalization in a decentralized environment compared to FL.

\begin{figure*}
 \begin{subfigure}{0.32\textwidth}
     \includegraphics[width=1\textwidth]{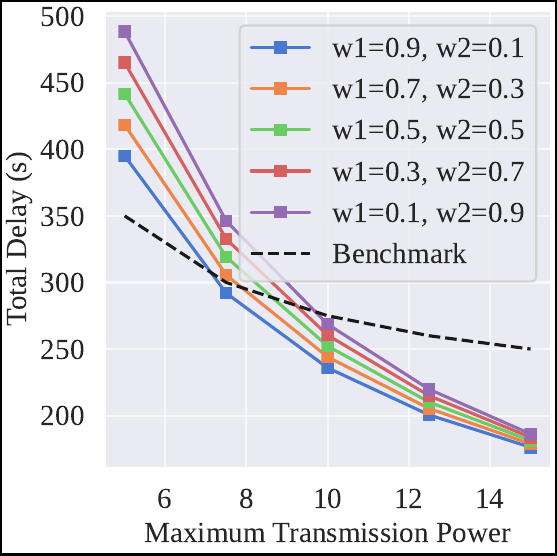}
     \caption{Total delay versus maximum transmit power}
     \label{fig:delay_vs_power}
 \end{subfigure}
 \hfill
 \begin{subfigure}{0.32\textwidth}
     \includegraphics[width=1\textwidth]{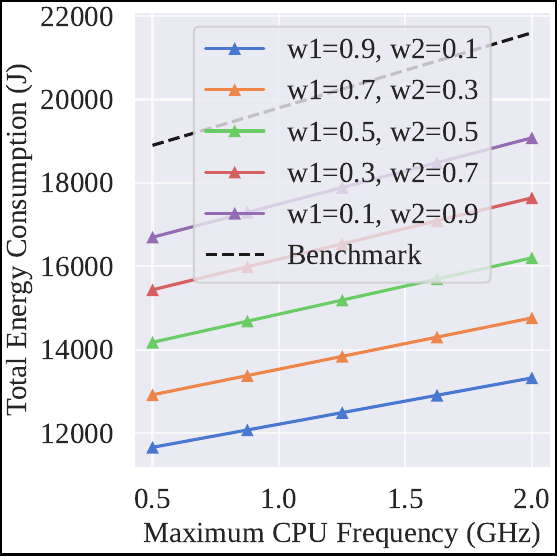}
     \caption{Total energy consumption versus maximum CPU frequency}
     \label{fig:energy_vs_cpu}
 \end{subfigure}
 \hfill
 \begin{subfigure}{0.32\textwidth}
     \includegraphics[width=1\textwidth]{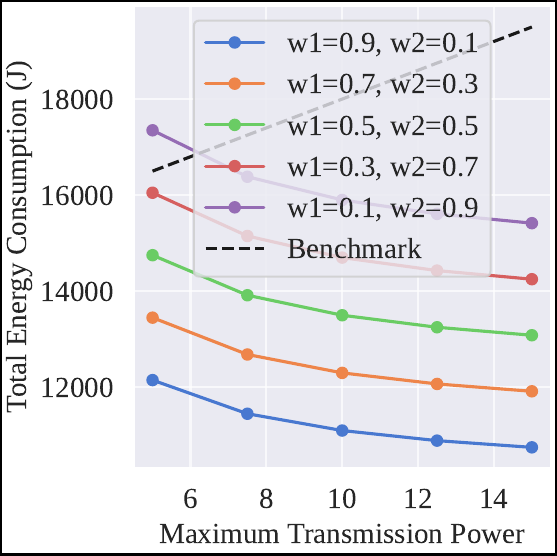}
     \caption{Total energy consumption versus maximum transmit power}
     \label{fig:energy_vs_power}
 \end{subfigure}
 \caption{Energy–latency trade-offs under different system constraints and weight settings $(w_1,w_2)$.}
 \label{fig:epoch_comp}
\end{figure*}

\subsection{Experiments on Client Communication Efficiency and Computational Load}
To compare the impact of the proposed split distributed framework over the federated framework based on computation time and efficiency, we have also measured the time taken in seconds and the computation power units taken in TFLOPS per forward and backward pass for each epoch. From figures \ref{fig:time_loss}, \ref{fig:time_ac}, and \ref{fig:time_f1}, we can see SpL-4-1 demands much lesser client processing times for forward and backward passes compared to federated models for training for an equivalent number of epochs. The average forward and backward pass processing time taken by clients for equivalent epochs of training for SpL-4-1 comes out to 5.85 seconds, compared to federated models, which reach almost 7 seconds. Figures \ref{fig:tflop_loss}, \ref{fig:tflop_acc}, \ref{fig:tflop_f1} also show that the computation power required for the convergence from client side is also much lesser for SpL-4-1 with a total of 72.70 TFLOPs for all clients (18.175 TFLOPs per client) compared to federated models which takes a significantly higher total of 109.87 TFLOPs from all the clients (27.48 TFLOPs per client), as shown in Table \ref{tab:time_tflops_comparison}. From a technical standpoint, the rapid convergence of SL can be attributed to the fact that only a portion of the model is trained on edge devices while the rest remains on the server. 
Unlike FL, where the entire model is trained locally before aggregation, SL ensures that intermediate representations are optimized centrally, leading to more stable gradient updates and improved learning efficiency. This better performance and computational efficiency of the split learning algorithm was also noted in previous works \cite{vepakomma2018split,singh2019detailed,10572484}.

\subsection{Experiments with cut-layer positions}
To analyze the impact of the model partitioning strategy on training dynamics, we experimented with three different cut layer positions: 4, 8, and 12, corresponding to the end of the first, second, and third convolutional blocks, respectively, as explained in the model architecture in a previous discussion. The results, visualized in Figure \ref{fig:cut_exp_comp}, reveal a critical trade-off between client computational load, communication time, and convergence speeds. Figures \ref{fig:cut_acc_epoch} and \ref{fig:cut_f1_epoch} show that the cut layer position doesn't impact the performance much. The analysis of processing time (Figures \ref{fig:cut_acc_time} and \ref{fig:cut_f1_time}) also reveals similar trends with the heaviest client-side model (position 12) taking longer processing time to converge.
In comparison, the lightest client model (position 4) takes significantly less time. This phenomenon can be attributed to communication overhead: splitting late at position 12 requires transmitting a large, high-dimensional activation tensor from the client, whereas the tensor from position 4 is more spatially compressed, reducing the data transmission bottleneck in each round. Figures \ref{fig:cut_acc_tflop} and \ref{fig:cut_f1_tflop} clearly illustrate the relationship between the split position and the client's computational load, measured in TFLOPs. An earlier split at position 4 requires the least computational effort from the client, achieving high performance with approximately 20 TFLOPs, and conversely, splitting at position 12 places the heaviest load on the client, demanding more than 60 TFLOPs to reach convergence. This directly confirms that offloading more layers to the server reduces the client's processing requirements. From a technical standpoint, the choice of the cut layer presents a fundamental trade-off. An early split (e.g., position 4) is ideal for resource-constrained edge devices with limited computational power. However, this comes at the cost of longer processing times at the server. A later split (e.g., position 12) accelerates training in terms of the lesser time required by the server to complete the training process for each client. However, it requires clients with greater computational capacities. Therefore, the optimal cut position is application-dependent, necessitating a balance between the hardware capabilities of client devices and the desired training speed.

\subsection{Experiments with number of clients}
In Figure \ref{client_comp}, we have visualized the results for our experiments on training the SL model on different numbers of clients. Specifically, we have shown results for training with client counts 2, 4, 6, 8, 12, and 16. Figures \ref{fig:client_acc_epoch}, \ref{fig:client_acc_time}, and \ref{fig:client_acc_tflop} show accuracy scores across the different client counts with respect to epochs, average time, and total TFLOPs, respectively. Figures \ref{fig:client_f1_epoch}, \ref{fig:client_f1_time}, and \ref{fig:client_f1_tflop} show the F1-scores across the different client counts with respect to epochs, average time, and total TFLOPs. These figures show that increasing the number of clients leads to faster convergence in training for the SL framework. At the same time, the model performance remains approximately the same with a few deviations. 
The communication overhead for FL increases exponentially with the number of participating clients due to full model weight transmission during each round.
In contrast, with SL, we only require transmitting the activations and gradients of a partial model, significantly lowering bandwidth requirements. 

\subsection{Numerical Validation of the Joint Optimization Framework}

 In Figure \ref{fig:delay_vs_power}, we plot the total latency against the maximum power budget. As expected, increasing the transmit power reduces communication delays. This effect is particularly pronounced when the system prioritizes latency minimization ($w_2$ dominant). Figure \ref{fig:energy_vs_cpu} investigates the impact of maximum CPU frequency. While higher frequencies reduce computation time, they also incur quadratic energy penalties, reflecting the trade-off between computational efficiency and speed. The figure illustrates how the weighting parameter $(w_1,w_2)$ shifts the balance between these two effects. Figure \ref{fig:energy_vs_power} shows the total energy consumption as a function of the maximum transmit power in normalized units. The proposed allocation strategy shows a consistently lower energy consumption compared with the benchmark scheme, and the gap widens for more energy-focused weight settings.

\section{Conclusions and Future Work}\label{conclusion}

In this work, we proposed a split learning framework for IoMT malware detection using image-based representations. The framework strategically distributes computation between edge devices and servers, achieving improved performance in terms of accuracy, convergence speed, and computational efficiency compared to conventional federated learning. Experimental results demonstrate that the proposed approach not only enhances malware detection accuracy but also reduces client-side computation and communication overhead, making it highly suitable for resource-constrained IoMT environments. While the presented framework addresses critical challenges, several avenues remain open for exploration. Future research can extend this work by investigating transfer learning approaches (leveraging pre-trained models to reduce training cost on IoMT devices), vertical split learning (to improve privacy guarantees in multi-party collaborations), and adversarial robustness techniques (to enhance resilience against gradient leakage and poisoning attacks). These directions will further strengthen the applicability of split learning for real-world IoMT security.

\section{Acknowledgment}
\thanks{This work was supported by CHANAKYA Fellowship Program of TIH Foundation for IoT \& IoE (TIH-IoT) received by Dr. Vinay Chamola under Project Grant File CFP/2022/027 and an ARC Discovery Project. }

\bibliography{bibliography.bib}
\bibliographystyle{IEEEtran}

\end{document}